\documentclass[%
reprint,
superscriptaddress,
 amsmath,amssymb,
 aps,
floatfix,
intlimits,
]{revtex4-2}

\usepackage{braket}
\usepackage{siunitx}
\usepackage{nicefrac}
\usepackage{graphicx}
\usepackage{dcolumn}
\usepackage{bm}
\usepackage{xcolor}
\usepackage[makeroom]{cancel}
\usepackage{mathtools}
\usepackage{comment}

\usepackage{enumitem}
\usepackage{interval}
\usepackage{amsmath,amssymb}
\usepackage{interval}
\usepackage[flushleft]{threeparttable}
\usepackage{booktabs}
\usepackage{siunitx}
\usepackage{textgreek}
\usepackage{lineno}
\DeclareSIUnit\barn{b}
\DeclareSIUnit\year{y}

\begin{document}

\preprint{APS/123-QED}

\title{Revisiting the excitation of the low-lying \textsuperscript{181m}Ta isomer in optical laser-generated plasma}

\author{Simone Gargiulo}
\email{simonegargiulo2@gmail.com}
\affiliation{%
 Institute of Physics (IPHYS), Laboratory for Ultrafast Microscopy and Electron Scattering (LUMES), École Polytechnique Fédérale de Lausanne (EPFL), Lausanne 1015 CH, Switzerland.\unpenalty~}%
\author{Ivan Madan}%
\affiliation{%
 Institute of Physics (IPHYS), Laboratory for Ultrafast Microscopy and Electron Scattering (LUMES), École Polytechnique Fédérale de Lausanne (EPFL), Lausanne 1015 CH, Switzerland.\unpenalty~}%
\author{Benoit Truc}%
\affiliation{%
 Institute of Physics (IPHYS), Laboratory for Ultrafast Microscopy and Electron Scattering (LUMES), École Polytechnique Fédérale de Lausanne (EPFL), Lausanne 1015 CH, Switzerland.\unpenalty~}%
\author{Paolo Usai}%
\affiliation{%
 Institute of Physics (IPHYS), Laboratory for Ultrafast Microscopy and Electron Scattering (LUMES), École Polytechnique Fédérale de Lausanne (EPFL), Lausanne 1015 CH, Switzerland.\unpenalty~}%
 \author{Kjeld Beeks}
\affiliation{%
 Institute of Physics (IPHYS), Laboratory for Ultrafast Microscopy and Electron Scattering (LUMES), École Polytechnique Fédérale de Lausanne (EPFL), Lausanne 1015 CH, Switzerland.\unpenalty~}%
\author{Veronica Leccese}
\affiliation{%
 Institute of Physics (IPHYS), Laboratory for Ultrafast Microscopy and Electron Scattering (LUMES), École Polytechnique Fédérale de Lausanne (EPFL), Lausanne 1015 CH, Switzerland.\unpenalty~}%

\author{Giovanni Maria Vanacore}%
\affiliation{%
 Institute of Physics (IPHYS), Laboratory for Ultrafast Microscopy and Electron Scattering (LUMES), École Polytechnique Fédérale de Lausanne (EPFL), Lausanne 1015 CH, Switzerland.\unpenalty~}%
\affiliation{Current address: Department of Materials Science, University of Milano-Bicocca, Milano 20125 IT, Italy.\unpenalty~}
\author{Fabrizio Carbone}
\email{fabrizio.carbone@epfl.ch}
\affiliation{%
 Institute of Physics (IPHYS), Laboratory for Ultrafast Microscopy and Electron Scattering (LUMES), École Polytechnique Fédérale de Lausanne (EPFL), Lausanne 1015 CH, Switzerland.\unpenalty~}%


\begin{abstract}
The excitation of the \textsuperscript{181m}Ta isomer in the laser-plasma scenario was claimed to have been observed more than two decades ago. However, the reported experimental findings -- and the respective high excitation rate -- were later questioned as they could not be reproduced theoretically. The controversy has remained open ever since. In this work, we reinvestigate both theoretically and experimentally the \textsuperscript{181m}Ta nuclear excitation in an optical laser-generated plasma. Experimentally we have found no evidence for such an excitation process as consistently predicted by previous and our theoretical models. 
\end{abstract}     

\maketitle
Since their discovery in 1921~\cite{hahn1921neue}, isomers have played a pivotal role in many domains of physics, encompassing fundamental research to practical applications~\cite{walker2005ups,dracoulis2016review}. These long-lived nuclear-excited states have held promise for the realization of compact energy storage for several decades~\cite{roberts1997importance,aprahamian2005long,carroll2007nuclear}. However, a process that could efficiently enable their activation or depletion has yet to be discovered. This challenge has recently nurtured the study of electromagnetic processes of nuclear excitation~\cite{harston1999mechanisms,feng2022femtosecond}, particularly focusing on those involving an interaction with the atomic surrouding~\cite{dzyublik2013general,berengut2018resonant,chiara2018isomer, wu2019mo, rzadkiewicz2021novel,bilous2020electronic, wu2022dynamical, gargiulo2022nuclear_NEEC_EXI, gargiulo2022nuclear, qi2023isomeric}. These processes generally demand specific conditions: for instance, the presence of photons, free electrons, and vacancies, as well as precise energy-matching between the nuclear levels and the energy of the incoming particle. In this context, plasma emerges as a unique environment, where many of these needs may be fulfilled at once. Such plasma can be generated when a high-intensity femtosecond optical laser pulse interacts with matter. This interaction initiates a cascade of physical processes in the target material, including ionization, refraction, thermal and hydrodynamic expansion. A complete description of all the relevant nuclear excitation mechanisms occurring within this chaotic setting posed a serious challenge from the beginning~\cite{doolen1978nuclear_PRL, harston1999mechanisms, gosselin2004enhanced}. The importance of this modeling gains even more significance when considering the implications these processes may have for isomer populations in astrophysical plasma. As isomers do not share the same neutron capture cross-section with their ground state, their presence could impact nucleosynthesis in stellar sites~\cite{carroll1989accelerated,schumann1998coulomb,helmrich2014coupling}, influencing the abundance of elements across the Universe. 

The simplest electromagnetic process of nuclear excitation one may think of -- under such circumstances -- is the direct absorption of resonant photons. Direct photoexcitation of an isomer -- activated directly from the ground state with an X-ray light source -- was first shown in 1978 in a $^{57}$Fe target~\cite{cohen1978nuclear}.

This successful observation was followed by several attempts to induce excitations in \textsuperscript{235}U -- trying to exploit also other excitation mechanisms -- in a laser-generated plasma scenario. In 1979, the excitation of the \SI{76}{\electronvolt} isomer in \textsuperscript{235}U plasma was 
 reported~\cite{izawa1979production}, but a subsequent duplication of the experiment in 1991 did not find such evidence~\cite{arutyunyan1991cross} and the discrepancy remained unresolved.

In 1999, a claim was made involving the photoexcitation of \textsuperscript{181m}Ta in a laser-plasma environment~\cite{andreev1999excitation}, which was followed by Ref.~\cite{andreev2000excitation}.
In these works, a reference target made of tungsten (W) was used to assess the occurrence of the isomer activation in \textsuperscript{181}Ta.
 W provides a very different nuclear excitation spectrum compared to \textsuperscript{181m}Ta while offering similar plasma properties, being the immediate consecutive element on the periodic table.  The composition of a W sample generally includes different isotopes according to their abundance. The first excited nuclear states of the most abundant W isotopes are in the range $\SI{46.48}{}-\SI{100}{\kilo\electronvolt}$ and half-life $T_{1/2}\simeq\SI{0.185}{\nano\second}-\SI{1}{\nano\second}$ and connected to the ground state through E2 or M1 transitions. By contrast, the first isomeric level ($\mathrm{J^\pi=9/2^-}$) of \textsuperscript{181m}Ta is characterized by an energy $E_\mathrm{n}=\SI{6.237}{\kilo\electronvolt}$ and a half-life $T_{1/2}=\SI{6.05}{\micro\second}$~\cite{ENSDFdatabase} and connected to the ground state ($\mathrm{J^\pi=7/2^+}$) through an E1 transition.  Consequently, we expect the nuclear decay profiles from the two targets to differ significantly, at the very least in their temporal characteristics, even considering the unlikely event of excitation of any nuclear level of W in a plasma with a temperature of a few keV.

Despite this, the time dependence of the signals originating from the laser-generated plasma measured in Refs.~\cite{andreev1999excitation,andreev2000excitation} is fairly similar for both targets. This observation suggests a response primarily dominated by the detector afterglow or by the ion-wall interactions,  rather than nuclear \textgamma-decays. Nevertheless, in these studies, the excess counts in the Ta experiment, relative to W, were interpreted as the signature of the \textsuperscript{181m}Ta isomer deexcitation. With this attribution, the authors suggested an excitation from \SI{e2}{} to \SI{e3}{} nuclei per laser shot at intensities ranging from \SI[print-unity-mantissa=false]{1e16}{} to \SI{5e16}{\watt\per\square\centi\meter}. These experiments were later questioned~\cite{gobet2008particle,gobet2011nuclear} as theoretical models could not explain such a high excitation rate, leaving the controversy open. A 2017 study~\cite{savel2017direct}, which relied on the observation of delayed electrons, estimated a lower number of \SIrange[range-phrase = --, range-units = single]{30}{50}{} nuclei excited per laser shot. However, this measurement again relied on the excess counts (of Ta compared to W), and the excitation rate was still theoretically unexplainable~\cite{tkalya2012cross}. 

Because laser-induced plasma is a promising route for the investigation of low-lying nuclear levels and since the reported results could not be rationalized by theory, in this Letter, we replicated the experiment with an improved setup. Our objectives were twofold: to challenge (i) the existing theoretical model with an experimental test and (ii) the initial observation with an improved experimental setup. In this Letter, we show that the excitation of the low-lying isomeric state of \textsuperscript{181}Ta was not possible in the previous literature. 

Regardless of the excitation mechanism taking place in the plasma, one could imagine the following events to occur. Initially, the plasma is formed over a timescale comparable to that of the laser pulse. This plasma will be characterized by highly charged ions, free electrons, and a photon distribution. At this time, many radiative and nonradiative atomic processes may take place, such as radiative recombination or dielectronic recombination~\cite{mitchell1983dielectronic,ali1990dielectronic}. Simultaneously, nuclei can be excited by the direct absorption of photons or through other processes, namely nuclear excitation by inelastic electron scattering (NEIES), nuclear excitation by electron capture (NEEC), nuclear excitation by electron transition (NEET), and many others. These processes require the presence of electron or photon distributions with energies comparable with the nuclear level targeted for excitation. Therefore, these processes can be sustained as long as the plasma is significantly hot, i.e. over a timescale comparable with the plasma lifetime. For laser intensities on the order of \SI[print-unity-mantissa=false]{1e16}{\watt\per\square\centi\meter}, this timescale extends to a few tens of picoseconds. It is within this time window that plasma radiation arising from bremsstrahlung and radiative recombinations can be expected.  Simultaneously, nuclei may undergo excitation through one of the processes mentioned earlier. If this happens, a nuclear decay follows involving \textgamma-radiation and internal conversion. These decays occur according to the nuclear level half-life $T_{1/2}$, when the atom is in a neutral state, or longer if ionized, as the IC channel can be partially or totally suppressed. When the effective nuclear half-life is longer than the plasma lifetime, time-resolved spectroscopic techniques can distinguish between plasma radiation and nuclear signals. 

\begin{figure}[!htbp]
\centering 
\includegraphics[width=\linewidth]{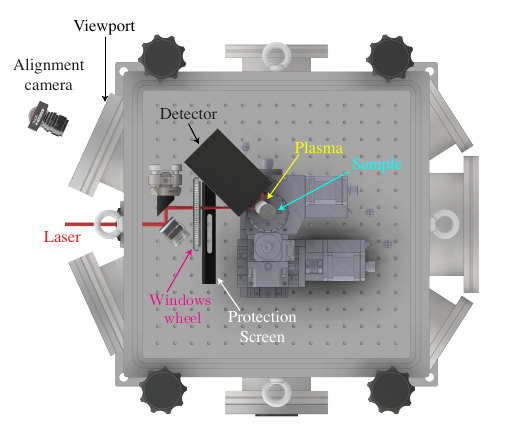} 
\caption{\textbf{3D rendering of the experimental setup designed for the observation of nuclear decays in a laser-plasma scenario.}}
\label{fig:setup} 
\end{figure}

Our experiment, shown in Fig.~\ref{fig:setup}, has been designed to enable such discrimination. The laser beam enters the chamber and then is focused on the target through a parabolic mirror, with a focal length of \SI{10}{\centi\meter}. 
\noindent
The surface of the sample is continuously moved (rotating the sample surface by $\sim \SI{0.5}{\micro\meter}$ per laser pulse and then shifting vertically after one cycle is completed) through a four axes mechanical stage. A deposit screen located below the detector (not visible in Fig.~\ref{fig:setup}) stops and collects the expanding plasma. The details of the screen and its positioning with respect to the detector are described in Ref.~\cite{gargiulo2023electromagnetic}. The detector is positioned directly above the deposit screen, facing downward to prevent any deposition on its surface. This deposit screen is needed for the detection of nuclear decay with $T_{1/2}\sim\SI{}{\micro\second}$; otherwise, ions will fly beyond the detector before their nuclear decay, lowering the detection solid angle. The rotating windows wheel -- which incorporates sapphire windows -- and the aluminum protection screen prevent deposition on the optics and allow a quick replacement of the window.
Table~\ref{tab:Laser} resumes the characteristics of the laser pulse as it arrives on the sample. 

\begin{table}[!hbtp]
\centering
\caption[Characteristics of the laser pulse.]{\label{tab:Laser} 
\textbf{Characteristics of the laser pulse.} $E_{\mathrm{pulse}}$ is the energy per pulse, $f_{\mathrm{rep}}$ is the repetition rate, $\lambda_\mathrm{l}$ is the laser wavelength, while $D_{\rm focal}$ is the FWHM of the laser spot at the focus of the parabolic mirror measured with a beam profiler. $\tau_\mathrm{pulse}$ is the pulse duration at the chamber entrance and $\lambda$ the laser wavelength. $I_\mathrm{peak}$ is the laser peak intensity.
} 

\begin{tabular}{S|S} \toprule
{$E_{\mathrm{pulse}}$} & {$\SI{1.25}{\milli\joule}$}\\[0.3cm]
{$f_{\mathrm{rep}}$} & {$\SI{4}{\kilo\hertz}$}\\[0.3cm]
{$D_\mathrm{focal}$} & {$\SI{10}{\micro\meter}$}\\[0.3cm]
{$\tau_\mathrm{pulse}$}&{\SI{40}{\femto\second}}\\[0.3cm]
{$\lambda_l$} & {$\SI{0.78}{\micro\meter}$}\\[0.3cm]
{$I_\mathrm{peak}=$} & {$\SI{4e16}{\watt\per\square\centi\meter}$}\\
\bottomrule
\end{tabular}
\end{table}
\noindent

The X-ray detector used is an AdvaPix TPX3~\cite{advapix}, which has time and energy resolutions of \SI{1.56}{\nano\second} and \SI{1}{\kilo\electronvolt}, respectively. The detector presents a hardware threshold of $D_\mathrm{th}=\SI{3}{\kilo\electronvolt}$; if the energy released in a single pixel is smaller than $D_\mathrm{th}$ the charge is lost and the event undetected.  The CCD of the detector is protected from deposition by a \SI{25}{\micro\meter} Kapton foil: this makes the detector sensitive only to the radiative \textgamma-decay and not to converted electrons. While plasma emission can be considered instantaneous compared to the detector resolution, the measured X-ray radiation is broadened in time by charge transport over the chip and electronics response. In our measurements, this broadening has a FWHM $\sim \SI{50}{\nano\second}$ and limits the minimum detectable isomer decay lifetime to $T_{1/2}\gg\SI{50}{\nano\second}$. More details about the setup design and characterization can be found in Ref.~\cite{gargiulo2023electromagnetic}. 

\begin{figure*}[!hbtp]
\centering 
\includegraphics[width=\linewidth]{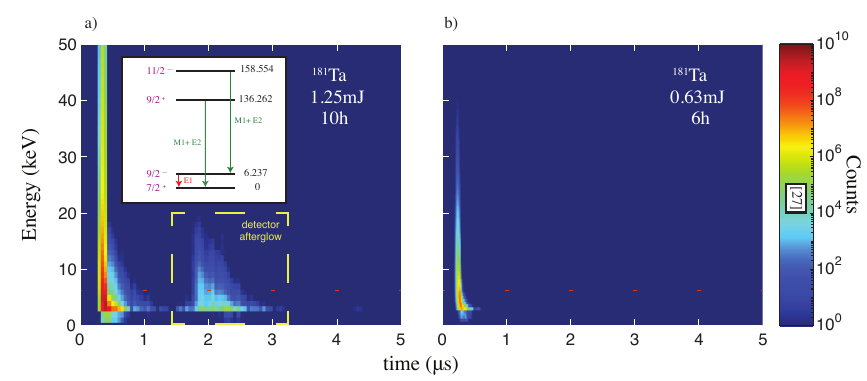} 
\caption{\textbf{Optical-laser generated plasma experiment: 2D energy-time map.} \textbf{a)} Experiment performed with $E_\mathrm{pulse}=\SI{1.25}{\milli\joule}$ and integrated for 10 hours. The time bin has been chosen to be \SI{50}{\nano\second}, while the energy bin is \SI{1}{\kilo\electronvolt}. The total number of events in this image is approximately equal to \SI{6.2e9}{}. Accounting for detector efficiency and the absorption of the Kapton foil used to protect the CCD, counts rise up to \SI{9e9}{}. The red dashed line indicates the energy position at which it would have been expected to see the nuclear decay of the first isomeric state of \textsuperscript{181}Ta. The color axis starts at $10^0$ to improve readability. However, with this choice, energy-time bins with a single count are indistinguishable from those with zero counts. In the Supplemental Material~\cite{SM}, the same histogram is replicated using a different colormap and a lower limit for the color axis. 
The white box on the color bar represents the count range ($10^4$--$10^5$) where, according to the excitation rate reported in Ref.~\cite{andreev1999excitation}, an isomer depletion signal was anticipated. Based on theoretical calculations, the expected counts range from \SI{e-9}{} (Bremsstrahlung) to \SI{e-1}{} (TDE) for the same integration period, which aligns with the absence of observed isomer depletion. In the inset, the level scheme of \textsuperscript{181}Ta with the first three excited nuclear states. Energies are expressed in keV. \textbf{b)} Experiment performed with $E_\mathrm{pulse}=\SI{0.63}{\milli\joule}$ and integrated for 6 hours. The total number of events equals \SI{2.83e9}{}.}
\label{fig:2D_histo} 
\end{figure*}

Fig.~\ref{fig:2D_histo}a reports the outcome of our laser-plasma experiment using a \textsuperscript{181}Ta cylindrical target. This energy-time map was acquired over an integration time of approximately \SI{10}{\hour} ($\SI{3.48e4}{\second}$), merging 116-time series of \SI{300}{\second} each and is reconstructed applying the built-in pixel-dependent energy calibration. 

Distinct features are noticeable. Initially, there is a broad, prompt, and intense signal at the start of the measurement window, corresponding to the plasma ignition. 
Subsequently, a weaker delayed signal emerges after $\sim \SI{1.5}{\micro\second}$. To understand the origin of this delayed signal, we conducted an analysis that focused on the specific pixel positions from which these counts originated. We determined that this secondary signal is an artificial detector afterglow, resulting from the detector's exposure to high radiation flux and its slow electronic response. It is important to note that the detector's readout speed of 38 million counts per second is based on conditions of uniform and continuous illumination. As our radiation signal is mainly concentrated on a few hundred of $\SI{}{\nano\second}$, the electronics is not fast enough to keep the pace. 
However, the occurrence of the afterglow does not impede the detector's ability to measure weaker signals within the relevant time frame. This has been confirmed through experiments involving the addition of a \textsuperscript{55}Fe radioactive source.
When we reduced the pulse energy ($E_\mathrm{pulse}$) by half, the detector's electronic was able to effectively manage the radiation, preventing the appearance of the afterglow. The resulting 2D histogram for this second experimental measure is shown in Fig.~\ref{fig:2D_histo}b and more details are given in the Supplemental Material~\cite{SM}. It is important to mention that even when the pulse energy is halved ($I_\mathrm{peak}=\SI{2e16}{\watt\per\square\centi\meter}$), we remain in the same laser peak intensity range (between \SI{1e16}{} and \SI{5e16}{\watt\per\square\centi\meter}) reported in Ref.~\cite{andreev1999excitation}.

In both cases, there is no trace of a signal compatible with the decay of the first isomeric level of \textsuperscript{181}Ta, which was expected where the dashed red line is drawn. In these integration intervals, considering the photoexcitation rate derived by Ref.~\cite{andreev1999excitation}, there should have been at least $\sim 10^4-10^5$ detectable events in our experiments~\footnote{To arrive at this estimate, we took into account various parameters: an excitation rate $\lambda$ $\sim \SI{e3}{}-\SI{e4}{\per\second}$, a number of ions $N_\mathrm{i} \sim \SI{e9}{}$, a repetition rate $f_\mathrm{rep}=\SI{4}{\kilo\hertz}$, our detection angle $\Omega\sim\SI{0.04}{\steradian}$, an integration time spanning 6 to 10 hours. Additionally, we considered the internal conversion coefficient $\alpha_\mathrm{IC}=70.5$ of the first isomeric level. The specified range for $\lambda$ ($\SI{e3}{}-\SI{e4}{\per\second}$) was determined based on a conservative estimate regarding the number of ions in the plasma of Ref.~\cite{andreev1999excitation}, which suggests $N_\mathrm{i}\sim\SI{e10}{}$.}. Therefore, the absence of a signal seems to confirm the doubts about their observation in Refs.~\cite{andreev1999excitation,andreev2000excitation,savel2017direct}.

We can now focus on the theoretical description of the experiment and ask whether these results are consistent with the models.
In this context, the number of excited nuclei per laser shot can be expressed as follows~\cite{wu2018tailoring,gunst2018nuclear}:
\begin{equation}
N_{\rm exc}= N_\mathrm{i}\ \lambda\ \tau_\mathrm{plasma}\, ,
\label{eq:DecayExcited}
\end{equation}
\noindent
where $N_\mathrm{i}$ is the total number of ions present in the plasma, $\lambda$ is the excitation rate, and $\tau_\mathrm{plasma}$ is the plasma lifetime. If we take direct photoabsorption as the reference excitation process, the photoexcitation rate can be defined as

\begin{equation}
\lambda_{\gamma}= \int \sigma_{\gamma}(E_\gamma)\ \phi_\gamma (E_\gamma,T_\mathrm{e},n_\mathrm{e})\ \mathrm{d}E_\gamma,
\end{equation}
\noindent
and it represents the integral of the product between the photoexcitation cross-section $\sigma_\gamma$ and the photon flux $\phi_\gamma$. $\phi_\gamma$ can be modeled considering the Bremsstrahlung at the origin of the photon gas in the plasma. This plasma is defined by its electron temperature $T_\mathrm{e}$ and electron density $n_\mathrm{e}$~\cite{harston1999mechanisms,pratt1977bremsstrahlung}. These key properties, together with its expansion dynamics, can be derived following the scaling laws presented in Ref.~\cite{gibbon1996short} and the hydrodynamic model in Ref.~\cite{krainov2002cluster}. An upper limit for $\phi_\gamma$ can be obtained by assuming thermodynamic equilibrium (TDE) between the electron and photon fluxes and adopting a Planck distribution for $\phi_\gamma$. Comprehensive modeling of the plasma, which employs the collisional-radiative code \textsc{flychk}~\cite{chung2005flychk} is detailed in the Supplemental Material~\cite{SM}, which includes Refs.~\cite{brunel1987not,bonnaud1991laser,gibbon1992collisionless,ping2008absorption,KineticPhotons, karpeshin19993,wong2008introductory,cengel2003practical,jakubek2009energy}.

An alternative excitation process involves nuclear excitation by inelastic electron scattering (NEIES)~\cite{tkalya2012cross,gobet2022expected,qi2023isomeric}, where the nucleus is excited through a free-free electronic transition. Accordingly, the excitation rate can be expressed as:

\begin{equation}
\lambda_\mathrm{NEIES}= \int \sigma_\mathrm{NEIES}(E_\mathrm{e})\ \phi_\mathrm{e} (E_\mathrm{e},T_\mathrm{e},n_\mathrm{e})\ \mathrm{d}E_\mathrm{e},
\end{equation}
\noindent
where $\sigma_\mathrm{NEIES}$ is the NEIES cross-section taken from Ref.~\cite{tkalya2012cross} and $\phi_\mathrm{e}$ is the electron flux.

The outcomes of the theoretical description of our experimental conditions in Table~\ref{tab:Laser} are reported in Table~\ref{tab:Excited_Nuclei_bremm}. These predictions include the number of excited nuclei $N_\mathrm{exc}$, nuclear deexcitation by \textgamma-decay $N_\mathrm{deexc}^\gamma$ and the excitation rate $\lambda$, considering direct photoexcitation process (both assuming Bremsstrahlung at the origin of the plasma emission process and TDE conditions to retrieve an upper limit) and NEIES.

\begin{table}[h!]
\centering
\begin{threeparttable}
\sisetup{table-format=2.3}
\begin{tabular}{lSS|S} \toprule
{}& \multicolumn{2}{c|}{Photons} & {Electrons}\\
{}&{Bremsstrahlung}&{TDE}&{NEIES}\\ \midrule
{$N_\mathrm{exc}$} & {$\SI{1e-12}{}$}& {$\SI{9.10e-5}{}$}&{$\SI{7.70e-6}{}$}\\[0.3cm]
{$N_\mathrm{deexc}^\mathrm{\gamma}$} & {$\SI{1.4e-14}{}$}& {$\SI{1.27e-6}{}$}&{$\SI{1.08e-7}{}$}\\[0.3cm]
{$\lambda$} & {$\SI{5.84e-11}{\per\second}$}& {$\SI{5.31e-3}{\per\second}$}&{$\SI{4.49e-4}{\per\second}$}\\

\bottomrule
\end{tabular}
\end{threeparttable}
\caption{\textbf{Number of excited nuclei, deexcitations, and photons in plasma, through the process of photoexcitation -- considering bremsstrahlung or thermal radiation (TDE) as the origin of the photons gas -- and NEIES.} This table reports the number of excited nuclei ($N_\mathrm{exc}$), the number of nuclei that deexcite through \textgamma-decay ($N_\mathrm{deexc}^\gamma$), and the excitation rate ($\lambda$). These values have been calculated per laser pulse.
}
\label{tab:Excited_Nuclei_bremm}
\end{table}
\noindent
The estimates presented in Table~\ref{tab:Excited_Nuclei_bremm} indicate that over an integration period of 10 hours, no \textsuperscript{181m}Ta \textgamma-deexcitation can be observed (the upper limit is 0.58 counts in TDE and the lower is \SI{6.4e-9}{} in Bremsstrahlung): these findings are consistent with the 2D map presented in Fig.~\ref{fig:2D_histo}a. Lower probabilities are naturally expected for the measurement shown in Fig.~\ref{fig:2D_histo}b.
We must highlight that for the plasma temperature achieved in our experiment, the indirect isomer population through the excitation of higher-lying levels (such as the \SI{482}{\kilo\electronvolt} level) is negligible. The excitation of such a level, for example, is lower by approximately 20 orders of magnitude compared to the excitation rates reported in Table~\ref{tab:Excited_Nuclei_bremm}.

On the other hand, Refs.~\cite{andreev1999excitation,andreev2000excitation,savel2017direct} have reported in their experiments $N_\mathrm{exc}\sim 10^{2}-10^{4}$ excited nuclei per laser shot. Considering $N_\mathrm{i}\sim 10^{9}$ and a plasma duration $\tau_\mathrm{plasma}\sim \SI{10}{\pico\second}$, in line with the models presented in Ref.~\cite{wu2018tailoring,gunst2018nuclear}, these $N_\mathrm{exc}$ values correspond to an excitation rate of $\lambda_\gamma\sim 10^4-10^6~\SI{}{\per\second}$. When applying the theoretical framework here presented to the scenario of Refs.~\cite{andreev1999excitation}, we derive $\lambda_\gamma = \SI{1.41e-2}{\per\second}$ under TDE conditions. Even by incorporating their estimation for the plasma volume (stated to be \SI{0.1}{\micro\meter}), we recover $\lambda_\gamma=\SI{8.42}{\per\second}$ in TDE settings. Consequently, even under the most favorable conditions -- that are hardly met in a real experiment -- the discrepancy between the expected and reported excitation rate spans between 4 and 8 orders of magnitude.

The nonobservation of the isomer decay in our experiment fully aligns with the theoretical model presented in the Supplemental Material~\cite{SM} and in Refs.~\cite{wu2018tailoring,gunst2018nuclear}, hopefully closing the controversy on the excitation of the \textsuperscript{181m}Ta in a plasma generated by a femtosecond laser pulse. 

The theoretical model also provides an estimate for the deexcitation efficiency ratio, defined as the ratio between the number of isomeric \textgamma-decays and the total number of photons emitted by the plasma over its lifetime in the energy range $\SI{3}{\kilo\electronvolt}\leq E \leq \SI{50}{\kilo\electronvolt}$. This efficiency is of $\nicefrac{N_\mathrm{deexc}^\gamma}{N_\mathrm{plasma}^\gamma}\sim10^{-18}-10^{-20}$. Hence, given that the total counts detected in Fig.~\ref{fig:2D_histo}a are $\sim \SI{9e9}{}$, we conclude that to witness a single nuclear deexcitation would necessitate extensive measurements over $\SI[print-unity-mantissa=false]{1e9}{\hour}\sim\SI[print-unity-mantissa=false]{1e5}{\year}$. Additionally, these counts -- as detailed in the Supplemental Material~\cite{SM} -- are consistent with the Bremsstrahlung model.

Overall, our experimental conditions are extremely similar to previous attempts, and the large discrepancy observed can not be only explained by the slight differences present, such as wavelength or pulse duration. The absence of an observable signal, supported by theoretical models, suggests that the \textsuperscript{181m}Ta isomer excitation was not achieved in the previous attempts present in the literature. 

The major issue is that the claims of Refs.~\cite{andreev1999excitation,andreev2000excitation,savel2017direct} rely exclusively on the detection of excess counts between Ta and W samples and that this excess is associated with the deexcitation of the \textsuperscript{181m}Ta isomer. Yet, these excess counts might be due to several causes that do not necessarily invoke the isomer excitation and subsequent \textgamma-decay. These reasons may include detector afterglow, local variations in the target surface properties, impurities, or overlaps from ions interacting with chamber walls in the time window where nuclear decays are expected, as stated by Refs.~\cite{andreev1999excitation,andreev2000excitation}.
Therefore, relying on the excess counts between the two targets as a signature of the nuclear decay could mistakenly lead to the conclusion that nuclear excitation was achieved during the experiment. However, this perspective might be misleading. While there may be excess counts, the essential time-energy signature indicative of the isomer decay remains absent, as depicted in Figs.~\ref{fig:2D_histo}a and \ref{fig:2D_histo}b.

In this Letter, we addressed the isomer excitation of \textsuperscript{181m}Ta in a laser-plasma scenario, both from a theoretical and experimental perspective. For this purpose, we have designed a tabletop experimental setup to study laser-matter interaction and incorporated time-resolved spectroscopic techniques to discriminate nuclear decay from plasma radiation. 
The absence of a clear signature of the processes leads us to conclude that the activation of \textsuperscript{181m}Ta could not have been achieved in previous laser-plasma experiments. This result is coherent with the current theoretical understanding.
Looking at the improvements necessary for such observation, one has to act on the variables appearing in Eq.~\ref{eq:DecayExcited}. Providing denser plasma with higher energy pulses will both act on $N_\mathrm{i}$ and $\phi_\gamma$, enhancing the number of excited nuclei $N_\mathrm{exc}$.  Nonetheless, it is currently necessary to overcome several orders of magnitude to have meaningful statistics over a reasonable time span, such as hours or days.

\bigskip
S.G. has designed the setup following inputs from I.M., G.M.V., and F.C.. F.C. supervised the project. S.G., I.M, and G.M.V. have performed the initial experiments that led to this design. S.G. and I.M. have assembled the setup and performed the experiments with the help of B.T. and K.B.. S.G. and I.M. have established the measurement and alignment procedure with inputs from P.U.. S.G. has performed the numerical estimates and calculations. S.G. has performed the data analysis. S.G. and V.L. have performed the measurement of the sample surface. The calibration of the detector using an \textsuperscript{55}Fe source was proposed by S.G. and performed by K.B. S.G. wrote the first draft of the manuscript. All the authors have contributed to the final version of the manuscript and discussed the outcomes of this research work. 
S.G., I.M. and F.C. acknowledge support from Google Inc. B.T. acknowledges support from ERC consolidator grant ISCQuM No 771346.

S.G. and F.C. would like to thank C. J. Chiara and A. P\'alffy for their useful feedback and comments.


\begin{thebibliography}{10}
\bibitem{SM}
``See {S}upplemental {M}aterial at [url will be inserted by publisher] for the details on the plasma hydrodynamic model, detector calibration with a \textsuperscript{55}{F}e radioactive source, and additional figures.''
\end{thebibliography}

\end{document}